\documentclass[twocolumn,secnumarabic,amssymb, nobibnotes, aps, prd]{revtex4}
\usepackage{amsmath}

\usepackage{color}
\usepackage{pstricks}
\usepackage{graphicx}
\usepackage{slashed}
\usepackage{epstopdf} 
\usepackage{mathtools}
\usepackage{braket}

\begin{document}

\title{On the Light dilaton in the Large $N$ Tri-critical O($N$) Model }

\author{Hamid Omid, Gordon W. Semenoff} 
\affiliation{ 
 Department of Physics and
Astronomy, University of British Columbia,
Vancouver, BC Canada V6T 1Z1 }

\author{L.C.R. Wijewardhana}

\affiliation{Department of Physics, University of Cincinnati, Cincinnati, Ohio 45220, USA}

\begin{abstract}
 
The leading order of the large  N limit of the O(N) symmetric phi-six theory in three dimensions has a phase which 
exhibits spontaneous breaking of scale symmetry accompanied by a massless dilaton which is 
a Goldstone boson. At the next-to-leading order in large N,  the phi-six coupling
has a beta function of order 1/N and it is expected that the dilaton
acquires a small mass, proportional to the beta function and the condensate. In this note, 
we show that this ``light dilaton''   is actually a tachyon.  This indicates an instability of the phase of 
the theory with spontaneously broken approximate scale invariance. 
 \end{abstract}

\maketitle

The scenario where  a quantum field theory can have a parametrically small beta function resulting in approximate scale invariance has attracted attention, particularly when the approximate
scale symmetry can be spontaneously broken, generating a pseudo-Goldstone boson in the form of a  light dilaton.  
For example,  the notion that the tree level
scale invariance of the $SU(2)\times U(1)$ electroweak theory is softly broken by the Higgs 
potential or dynamically broken by some physical mechanism
beyond the standard model leaves the Higgs boson
itself as the dilaton, with some testable physical consequences \cite{Goldberger:2008zz}-\cite{Antipin:2011aa}.   
Walking techni-color  \cite{Holdom:1984sk}-\cite{Cohen:1988sq} and confining and chiral symmetry
breaking  gauge field theories with approximate infrared conformal symmetry \cite{Appelquist:2010gy}
are also  scenarios where spontaneous breaking of approximate  scale invariance could play an
important role. 

\begin{figure}
\begin{center}
\includegraphics[scale=.4]{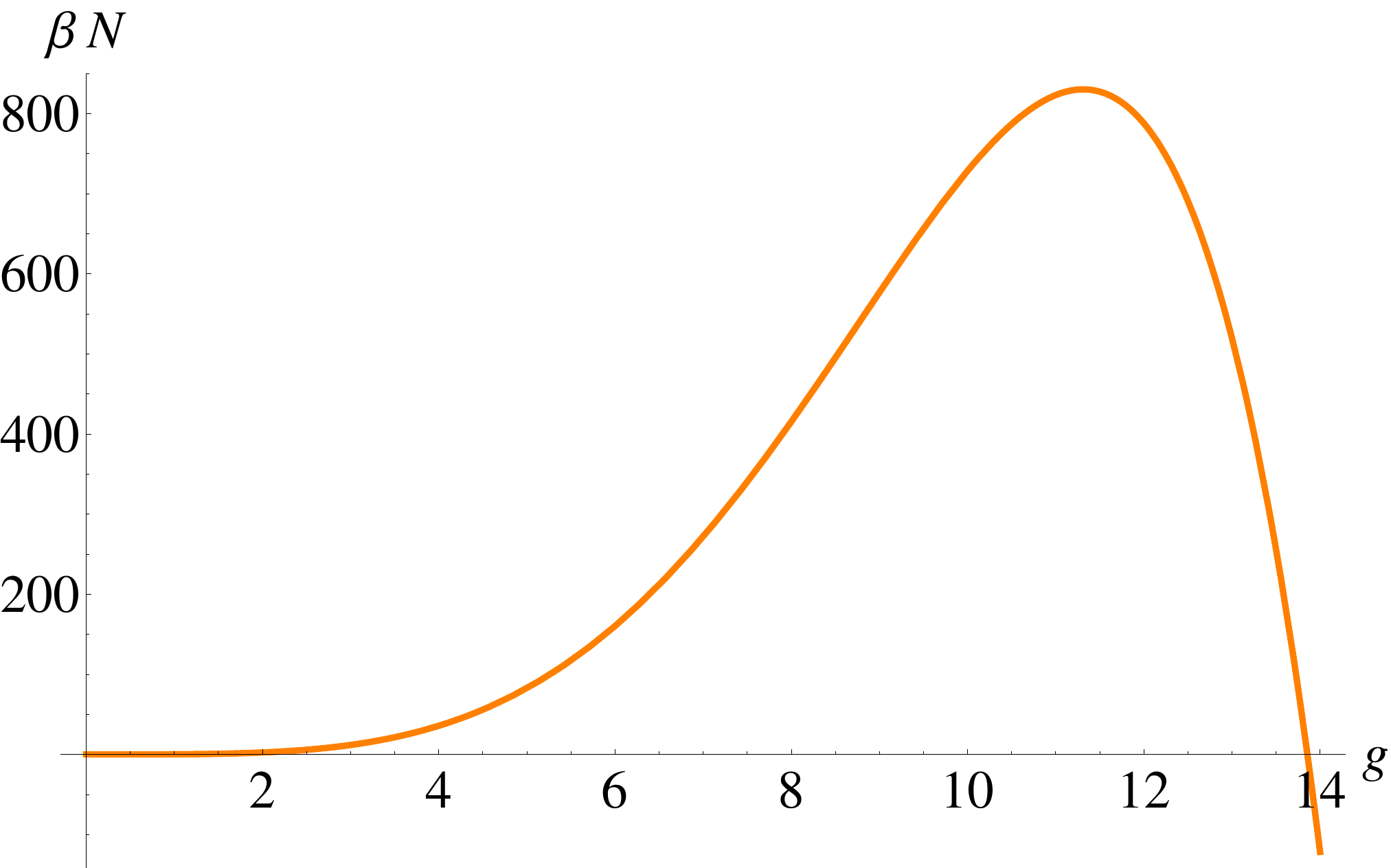}
 \caption{N$\times $ the beta function of large N regime of $g^2(\vec \phi^2)^3$ theory in three dimensions. The infrared fixed point is $g^2_{\rm IR}=0$ and the ultra-violet
 fixed point occurs at $g^2_{\rm UV}= {192}$.   The critical coupling where  in the infinite N limit scale symmetry breaking occurs is $g^2=(4\pi)^2\approx 158$.}
\label{beta}
\end{center}
\end{figure}

One of the prototypical examples of spontaneously broken scale symmetry occurs in the large N limit of the tri-critical O(N) symmetric
$ g^2(\vec\phi^2)^3$-theory in three space-time dimensions which 
is an interesting quantum field theory in its own right. 
The $g^2(\vec \phi^2)^3$ interaction is scale invariant at the classical level. 
Its beta-function is of order 1/N and it is therefore suppressed at large N.  As a result,   $g^2(\vec\phi^2)^3$ remains exactly marginal  at the leading order in the large N expansion and, at the next-to-leading order $g^2$, it becomes a slowly running coupling and the theory is approximately scale invariant. 
  The beta function, depicted in 
   FIG. \ref{beta},  has a trivial infrared fixed point at $g^2=0$.   In addition, as was argued long ago 
   \cite{Townsend:1975kh}-\cite{Appelquist:1982vd}, it exhibits a 
   nontrivial ultra-violet fixed point at $g^2=192$. The ultra-violet
fixed point renders  the  field theory   asymptotically safe in that the ultra-violet cut-off can  
be removed without forcing triviality \cite{Pisarski:1982vz}.

However, to complicate matters, Bardeen, Moshe and Bander \cite{Bardeen:1983rv} showed that, with sufficiently strong
coupling, the
  infinite N limit of $g^2(\vec \phi^2)^3$ theory 
    has a 
quantum phase transition to a phase where an O(N) singlet composite operator, $\frac{1}{\text{N}}\vec\phi^2$, gets an expectation value.
This operator has a non-zero scaling dimension and its expectation value  
breaks the  scale symmetry.  The phase transition occurs  at a critical value of the coupling, $g^{*2}=(4\pi)^2\approx 158$
 that is somewhat smaller than
the ultra-violet fixed point, $g^2_{\rm UV}= {192} $.  They also showed that the 
ultra-violet cut-off can be removed only when
$g^{2}$ is tuned to their fixed point,  $g^2\to g^{*2}$.   The condensate 
 breaks the exact scale symmetry of that limit and they showed that
the spectrum of the theory contains a massless dilaton. 
David, Kessler and Neuberger \cite{David:1984we,David:1985zz} 
did a careful analysis of the phase diagram of the infinite N model.  They also conjectured that, ``for all finite  $N$ the BMB 
(Bardeen-Moshe-Bander) 
phenomenon does not survive''.  The latter is something that we shall demonstrate to be so in the following.

 It was suggested in the original work of Bardeen,  Moshe and Bander \cite{Bardeen:1983rv} that,  if one considers
 the large, but not infinite, $N$ limit,  where the scale symmetry becomes approximate,  in their strong coupling phase, 
 the latter is represented as a spontaneously broken approximate symmetry and the dilaton becomes a quasi-Goldstone boson,
acquiring a mass of order 1/N.
 In this paper, we will examine this issue by studying  the
tri-critical O(N) vector model in the  leading and next-to-leading order of the large N 
limit. Of particular interest will be the interplay between the  loss of tune-ability of the coupling constant when it has
a renormalization group flow and dynamical breaking of scale invariance which 
is driven by strong coupling dynamics and occurs at a specific fixed point.  Our central conclusion will be that the
 ``light dilaton''  of this theory is actually a tachyon.  This indicates an instability of the phase of 
the theory with spontaneously broken approximate scale invariance.  
Our computation is in complete perturbative control, at least in the context of a renormalization group
improved large N expansion, when N is large enough.  We note that, potential instability, based on the fact
that this is ultimately a theory with a cubic potential was pointed out by Gudmindsdottir et.~al.~\cite{Gudmundsdottir:1985cp}.

 The composite operator effective action was originally computed 
by Townsend \cite{Townsend:1975kh} and our results are  in agreement with  his where they 
overlap, the main difference being that he studied O(N) symmetry breaking whereas we study the 
the massive phase which occurs near the tri-critical point. The O(N) vector field will be denoted $\vec\phi(x)$. 
We will find it convenient to describe the
theory by two variables,  the composite field $\chi (x)= \frac{1}{\text{N}}\vec\phi^2$ and an auxiliary field $M(x)$, whose expectation
value is proportional to the $\vec\phi$-field mass. Both $\chi(x)$ and $M(x)$ have classical dimension one and $\frac{M(x)}{\chi(x)}$ is dimensionless. 
Whenever $<M(x)>$ is not zero, the $\vec\phi$-field is massive and it does not obtain
an expectation value. 
We will find that, at the leading and next-to leading order in the 1/N expansion, the
 renormalized background field effective action  
 is
\begin{align}\label{veff_1}
S =\text{N}\int d^3x\left\{ \frac{\chi^3(x)}{6}  \left(g^2(M(x))-g^{*2}\left(\tfrac{M(x)}{\chi(x)}\right)\right)+~~ \right. \nonumber \\
\left. +\frac{ \partial M(x)\cdot\partial M(x)}{96\pi | M(x) |}+\ldots\right\}
\end{align}
where $g^2(M)$ is the running coupling at scale $M$ and 
$g^{*2}(x)$,  where $x=\frac{M}{\chi}$,  is the scale invariant part of the non-derivative terms in the effective
action, containing contributions of order one and of order $\frac{1}{N}$.  
The ellipses denote    contributions of  order $ \frac{1}{N^2}$ or higher of any type and terms with more
than two 
derivatives. 
Although $\chi$ is nominally a positive operator, an infinite normal ordering
constant has been subtracted from it so that it can now be either positive or negative.  The  couplings have been tuned so that terms 
proportional to  $(\vec\phi^2)^2$ or $\vec \phi^2$ are absent.  

To use the background field effective action (\ref{veff_1}) we should first solve the  equations 
which determine its extrema, 
\begin{align}
\frac{\delta S}{\delta\chi(x)}=0
~~,~~
\frac{\delta S}{\delta M(x)}=0
\label{eqmo}
\end{align}
Solutions of these equations are the classical fields which we shall denote by $M_0$ and $\chi_0$.   If there are more than one solution (there will
not be in our example), we should choose the solution where $S$, when evaluated on the solution, has the smallest real part. The expansion of the 
action in (\ref{veff_1}) in derivatives assumes that $M_0$ and $\chi_0$ are non-zero and that they are slowly varying functions, sufficiently so that the expansion in
their derivatives is accurate.    ($M_0$ and $\chi_0$ are usually
constants.) Then, in order to compute a one-particle irreducible correlation function of the fields $\chi(x)$ and $M(x)$, we take
functional derivatives of the
background field action $S$ by the variables $\chi(x)$ and $M(x)$, and we subsequently evaluate the resulting functions ``on-shell'' by
setting $\chi(x)$ and $M(x)$ to  $\chi_0$ and $M_0$,
respectively.  This  yields the renormalized, connected, 
one-particle-irreducible multi-point correlation functions of the quantum fields $\chi(x)$ and $M(x)$. 
For example, the connected two-point correlation functions
are found by inverting the one-particle irreducible two-point functions which are obtained as 
functional second derivatives of the effective action.  They are thus given by 
\begin{align}
\left[ \begin{matrix}  \left<\chi\chi\right>-\left<\chi\right>\left<\chi\right>&  \left<\chi M\right>-\left<\chi\right>\left<M\right>
\cr \left<M\chi\right>-\left<M\right>\left<\chi\right>&\left<MM\right>-\left<M\right>\left<M\right>
 \cr \end{matrix}\right]=\nonumber \\=
 \left. 
\left[ \begin{matrix}   \frac{\delta^2 S }{\delta\chi^2}&   \frac{\delta^2 S}{\delta\chi\partial M}
\cr  \frac{\delta^2 S}{\delta\chi\partial M} & \frac{\delta^2 S}{\delta M^2}
 \cr \end{matrix}\right]^{-1} \right|_{M,\chi=M_0,\chi_0}
\nonumber  \end{align}
For example, we obtain the composite operator correlation function
 \begin{align}
&\Braket{\tfrac{1}{\text{N}} \phi^2 (x)\tfrac{1}{\text{N}}\phi^2 (y)}-\Braket{\tfrac{1}{\text{N}} \phi^2 (x)}\Braket{\tfrac{1}{\text{N}}\phi^2 (y)}\nonumber \\
& =  \frac{1}{\text{N}}  \int\frac{d^3p}{(2\pi)^3}~e^{ip(x-y)}~  \frac{
48\pi\chi_0^2/M_0  
 }{p^2+
\frac{24\pi\chi_0^3 }{M_0}\beta(g^2(M_0))  }
\label{propagator1}
 \end{align}

Let us review a few interesting features of our results:
 \begin{enumerate}
 \item{}We are putatively working in the leading and next-to-leading orders of the large $N$ expansion. 
 The quantities in brackets in (\ref{veff_1}) are of order one and of order $\frac{1}{N}$.  The running coupling
 constant, $g^2(M)$, on the other hand, is the solution of the renormalization group equation using the beta function which
 is of order $\frac{1}{N}$.  If expanded in $\frac{1}{N}$, it contains all orders of $\frac{1}{N}$, multiplied by powers of logarithms of the mass scale
 ratio. This ``sum of leading logarithms'' is needed in order to accommodate possible very small or very large values
 of the condensate,  $M\sim \mu \exp\left( N\cdot \ldots\right)$. 
 \item{}At this order in the large N expansion, the only renormalization group function
 entering the affective action (\ref{veff_1}) is the running coupling constant $g^2(M)$ which is to be
 evaluated at the scale determined by the condensate. 
\item{} The generic features of the result in equation (\ref{propagator1}) do not depend much 
on the details of 
  the function $g^{*2}(x)$ in equation (\ref{veff_1}). It relies only on the 
 fact that its leading contribution
 at large N is independent of N and the fact that it is scale invariant, that is, it is a function of only   the dimensionless ratio 
 $\frac{M}{\chi}$ and $g^2$. Validity of the derivative expansion also requires 
  non-zero $\chi_0$ and $M_0$ as the classical solutions.  
 When evaluated on the solutions of the equations of motion (\ref{eqmo}), $g^{*2}=(4\pi)^2+{\mathcal O}
 (\frac{1}{\text{N}})$.  At leading order in large N, $g^{*}=4\pi$, the value of the coupling at the Bardeen-Moshe-Bander fixed point.
 \item{}When N goes to infinity, the beta function vanishes and $g^2(M)$ becomes $M$-independent and tuneable.   
 Consequently, at this limit, the action (\ref{veff_1}) has an extremum only when $g^2=g^{*2}$ where the potential is flat and does
 not determine the scale.  The fluctuations of the scale, $M$ or $\chi$, is in a flat direction and forms a massless dilaton.
 This is the Bardeen-Moshe-Bander solution.
 \item{}The signature of the dilaton is the presence of the pole in the correlation function of
  $\Braket{ \phi^2 (x)\phi^2 (y)}$  in equation (\ref{propagator1}).  Note that the mass of this pole is proportional to the beta function, $\beta(g^2(M))$.
  The latter vanishes at infinite $N$, leaving the dilaton massless in that limit.
  \item{} For large but finite N, the pole in the two-point function  (\ref{propagator1}) occurs at
  $$
  -p^2 = \frac{24\pi\chi_0^3 }{M_0}\beta(g^2(M_0)) 
  $$
The beta-function is positive,  $\beta(g^2(M_0)) >0$. However, 
 the values $\chi_0$ and $M_0$ which solve the equations of motion turn out to  
 have opposite signs. That opposite sign results in the mass squared in the pole in the propagator in (\ref{propagator1})
 having a negative sign
 and the dilaton has become  a tachyon.  The tachyonic mass indicates that the
 phase that we are describing unstable to fluctuations. 
   \end{enumerate}
 In the following, we will present our derivation of equation (\ref{veff_1}) and the simple computation leading
 from equation (\ref{veff_1}) to (\ref{propagator1}). 
  We consider the Euclidean quantum field theory which has N real scalar fields $\vec\phi=
(\phi^1,\phi^2,\ldots,\phi^N)$ and 
O(N) symmetry in three space-time dimensions. The classical Landau-Ginzburg potential is given by
\begin{align}\label{lgpot}
V(\vec\phi^2)= \frac{r}{2}\vec\phi^2+\frac{u}{4}(\vec\phi^2)^2+\frac{g^2}{6}(\vec\phi^2)^3
\end{align}
  \begin{figure}
\begin{center}
\includegraphics[scale=.45]{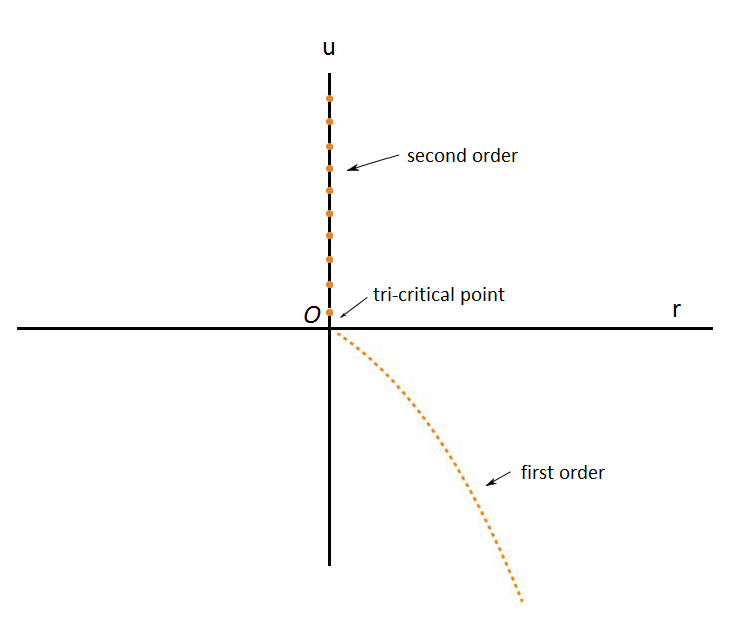}
 \caption{The phase diagram of the Landau-Ginzburg potential in equation (\ref{lgpot}). The tri-critical point $O$ appears at the intersection of
 a line of second order phase transitions and a line of first order phase transitions where the potential is equal to $\frac{g^2}{6}(\vec\phi^2)^3$.
  }
\label{figure1}
\end{center}
\end{figure} 
where 
$$
\vec\phi^2\equiv \sum_{a=1}^N \phi^a\phi^a
$$
When $u>0$, there is a line of second order phase transitions at $r=0$ as depicted in FIG. \ref{figure1}.
When $u<0$ there is a line of first order phase transitions.  These lines of transitions terminate at the 
tri-critical point $O$ where $u=r=0$.  At the classical level this phase structure persists for all positive values of $g^2$ and
the $g^2(\vec\phi^2)^3$  coupling   is exactly marginal. 

To examine fluctuations, we consider the Euclidean functional integral
\begin{equation}\label{pathintegral_1}
Z[j]=\int [d\vec \phi] ~e^{-\int d^3x L[\vec\phi,\vec j]}
\end{equation}
with  the Lagrangian density
\begin{equation}
L=\frac{1}{2}\partial_\mu\vec\phi\partial_\mu\vec\phi+\text{N}V( \vec\phi^2/\text{N}) - \vec j\cdot\vec\phi
\end{equation}
where $\mu=1,2,3$ and repeated indices are summed and we have introduced a source $\vec j(x)$ in
order to use the functional integral as a generating functional for correlators of $\vec\phi(x)$.  In order to study the large N limit,
we introduce two  auxiliary fields by inserting
\begin{align}
1&=\int_{-\infty}^{\infty} [d\chi(x)]\delta(\chi(x)-\vec\phi^2/\text{N})\\
&=\int_{-\infty}^{\infty} [d\chi(x)]\int_{-i\infty}^{i\infty}[dm^2(x)] e^{\int \frac{1}{2}m^2(\text{N}\chi-\vec\phi^2)}
\end{align}
into the functional integral (\ref{pathintegral_1}).  This introduces
two new fields $\chi(x)$ and $m^2(x)$ and it will allow us to integrate out the
scalar field $\vec\phi(x)$.  We must be careful to note that the integration for $m^2$ is on the imaginary
axis. We will find out and explicit form for the scale $M^2$ that was present in (\ref{veff_1}) as a function of $m^2$.
With these additional fields, the Lagrangian becomes
\begin{equation}
L=\frac{1}{2}\partial_\mu\vec \phi\partial_\mu\vec\phi+\frac{m^2}{2}\vec\phi^2-\text{N}\frac{m^2}{2}\chi+\text{N}V(\chi)-\vec j\cdot\vec\phi
\label{lagrangian}\end{equation} 
The $\vec \phi$-fields now appear in a quadratic form and
we integrate them exactly  to get an effective action
\begin{align}
S[m^2,\chi,\vec j]&=\frac{\text{N}}{2}{\rm Tr}\ln(-\partial^2+m^2)\nonumber \\
&+\int\left(\text{N} V(\chi) -\text{N} \frac{m^2}{2}\chi-\frac{1}{2}\vec j\frac{1}{-\partial^2+m^2}\vec j \right)
 \label{effectiveaction0}
\end{align}
To find the partition function, it remains to integrate $\chi$ and $m^2$,  
\begin{equation}\label{pathintegral}
Z[j]=\int [dm^2 d\chi ] ~e^{-S[m^2,\chi,\vec j]}
\end{equation}
This would yield a generating functional where functional derivatives with respect to $j$ give the correlation
functions of the $\vec\phi$-fields.

We will study the region of the phase diagram where the O(N) symmetry is not spontaneously
broken.  Instead, there will be a condensates   $\Braket{\chi(x)}$ and $\Braket{m^2(x)}$ which will result in a 
mass gap for the $\vec\phi$-field. 
To begin, it is
instructive  to put the source $j(x)$ to zero and to write the effective action in (\ref{effectiveaction0}) 
in an expansion in derivatives of the variable $\chi(x)$ and $m^2(x)$, 
 \begin{align}
&\frac{S}{\text{N}}=\int\left\{ \frac{\Lambda m^2}{4\pi^2}-\frac{|m|^3}{12\pi} 
+ V(\chi) -\frac{m^2\chi}{2} + \frac{\partial m.\partial m}{96\pi|m|}+\ldots\right\}
 \label{effectiveaction21}
\end{align}
where $\Lambda$ is the ultra-violet cut-off and the ellipses represent terms with more than two derivatives of $m$. 
We have dropped a constant term that is $m^2$ and $\chi$ independent. 
The effective action in (\ref{effectiveaction21}) has an ultra-violet divergent $\Lambda$-dependent term which must
be removed by renormalization. 
We can renormalize the expression  by introducing counter-terms. This is accomplished by replacing 
$V(\chi)$ by $V\left(\chi - \frac{\Lambda}{2\pi^2}\right)$.  Then,  after a field translation,  $\chi(x)
=\tilde\chi(x)+\frac{\Lambda}{2\pi^2}$, the cut-off dependent term cancels from (\ref{effectiveaction21}).  
Although $\chi(x)$ was originally a positive field, $\tilde\chi(x)$ can be positive or negative. We hereafter drop the tilde
from $\tilde\chi$. The second, third and fourth terms in (\ref{effectiveaction21}) are the effective potential for $m$ and $\chi$ at 
the leading order in the large N expansion. In the remainder  of this paper   we will choose the potential $V(\chi)$ to be 
the specific dimension-three operator
\begin{align}
V(\chi)=\frac{g^2}{6}\chi^3(x) +~{\rm counterterms}
\label{v}
\end{align}
where the counterterms will be needed to cancel divergences at higher orders in $\frac{1}{N}$.
With this choice, the field theory is scale invariant at the classical level and, since there are no logarithms in (\ref{effectiveaction21}),
it remains scale invariant at the quantum level in the leading order in the large N expansion. 

In the large N limit, we can use the saddle-point technique to evaluate the remaining functional integral (\ref{pathintegral}). 
The saddle points are field configurations which solve the equations of motion derived from the effective action (\ref{effectiveaction0}).
We will use the notation $\chi_0$ and $m_0$ to denote fields which satisfy the equations of motion. 
When the fields are constant, the saddle points are extrema of the renormalized effective potential obtained from (\ref{effectiveaction21}),
 The potential in (\ref{effectiveaction21}) has a line of extrema, located at $|m_0|=-4\pi \chi_0$  and 
these extrema exist only when the coupling constant  $g$ is set to the Bardeen-Moshe-Bander fixed point at $g \to g^*=4\pi$. 
To see this, consider the equation of motion for $m$.  This equation  does not involve the coupling constant.  It has the solution  
$$
|m_0|=\left\{ \begin{matrix} -4\pi \chi& \chi<0\cr 0& \chi>0\cr \end{matrix} \right.
$$
A massive solution exists only when $\chi$ is negative. Let us assume this is so. 
We can plug this solution into the effective action to get  (We use $\hat S$ to distinguish this partially on-shell action from $S$ in equation (\ref{veff11}) 
and elsewhere.)
\begin{align}\label{veff11}
\hat S =\text{N}\int \left\{- \frac{|\chi|^3}{6}  \left(g^2-(4\pi)^{2}\right) +\ldots\right\}
\end{align} 
and ask whether there is now a solution for $\chi$.  When $g > g^*$ this expression has no extrema and for $g < g^*$ there is no spontaneous symmetry breaking( $\chi=0$). However at $g=g^{*}$, the potential is flat and any (negative) constant $\chi_0$ is a solution.

To find the effective action to the next-to-leading order in large N, we use the background field technique. To implement this technique, we do the 
substitution
\begin{align}\label{contour}
\chi\to \chi+\delta\chi~,~m^2 \to m^2+i\delta m^2
\end{align}
and, following the recipe in \cite{Jackiw:1974}, we drop the linear terms in $\delta\chi$ and $\delta m^2$. Then, the
action expanded to quadratic order is
\begin{align}
&S=\frac{\text{N}}{2}{\rm Tr}\ln(-\partial^2+m^2)-\int \text{N}\frac{\Lambda m^2}{4\pi^2}\nonumber \\
&+\int\left(\text{N} V(\chi) -\text{N} \frac{m^2}{2}\chi-\frac{1}{2} j\frac{1}{-\partial^2+m^2}j \right)
 \nonumber  \\
&+\frac{\text{N}}{2}\int [\delta\chi, \delta m^2]\left[ \begin{matrix} \tilde V''(\chi)&  -i/2 \cr -i/2 & \Delta/2+{\mathcal J}[j]
 \cr \end{matrix}\right]
\left[ \begin{matrix} \delta\chi \cr  \delta m^2\cr\end{matrix}\right] 
+\ldots\label{effectiveaction1}
\end{align}
where  
\begin{align}
&\Delta(x,y) = \Bra{x}\frac{1}{-\partial^2+m^2}\Ket{y}\Bra{y}\frac{1}{-\partial^2+m^2}\Ket{x}
\label{Delta}
\\
&{\mathcal J}[x,y;j] =\frac{1}{\text{N}}\int dwdzj^a(w)j^a(z)\cdot \nonumber \\
&\cdot \Bra{w}\frac{1}{-\partial^2+m^2}\Ket{x} \Bra{x}\frac{1}{-\partial^2+m^2}\Ket{y}\Bra{y}\frac{1}{-\partial^2+m^2}\Ket{z}
\end{align}
When $m^2$ is a constant, 
\begin{eqnarray}
&\Delta(x,y)=\int \frac{d^3p}{(2\pi)^3}e^{ip(x-y)}\Delta(p) 
\nonumber \\
&\Delta(p) =\frac{1}{4\pi p}\arctan\frac{p}{2|m|} 
\label{delta}\end{eqnarray}

Before we proceed, we can use the action (\ref{effectiveaction1}) to study the spectrum of fluctuations in the 
infinite $N$ limit.  For this purpose, 
we invert the quadratic form in (\ref{effectiveaction1}) and find the 
propagator 
\begin{align} 
&\Braket{\tfrac{1}{\text{N}}\vec\phi^2~\tfrac{1}{\text{N}}\vec\phi^2}-\Braket{\tfrac{1}{\text{N}}\vec\phi^2}\Braket{\tfrac{1}{\text{N}}\vec\phi^2} = \Braket{\delta\chi~\delta\chi}
\nonumber  \\
&=\frac{\frac{2}{\text{N}}\Delta(p) }{1+2V''(\chi) \Delta(p)}  =
\frac{\frac{2}{\text{N}} \frac{ 1 }{ 4\pi p } \arctan \frac{ p }{ 2|m| }  }{1 -\frac{2m }{   p } \arctan \frac{ p }{ 2m}}\approx
\frac{3m}{\text{N}\pi}\frac{1}{ p^2}
\label{propagator}
 \end{align}
 where, in the last equality, we have put the condensate on shell and the coupling constant equal to the
 fixed point value, $4\pi$.  The last expression
reproduces the sum of bubble diagrams which would be expected from studying the Feynman diagrams for
this correlation function. The massless pole is due to the dilaton which is a Goldstone boson for spontaneous breaking of the scale  symmetry which is exact at this order in the large N expansion. 
 We can see that this massless pole is the only pole by studying  
  the denominator of (\ref{propagator}). 
\begin{align}
 1- \frac{2m}{ p}\arctan\frac{p}{2m}
 = \int_0^1 dx \frac{x^2} { \frac{4m^2}{p^2}+x^2  }
\end{align}
in the complex $-p^2$-plane. It is easiest to see from the 
  integral representation of the function  that the only zero  
 is at $-p^2=0$.   'There is also a cut singularity on the positive
 $-p^2$-axis beginning at $4m^2$ due to intermediate $\phi$-particle
 pairs. 
 
 To study the next order in the large N expansion, 
we do the Gaussian integral over the fluctuations in (\ref{effectiveaction1}) 
to get the effective action
\begin{align}
S=&\frac{\text{N}}{2}{\rm Tr}\ln(-\partial^2+m^2)\nonumber \\
&+\int\left(\text{N} V(\chi) -\text{N} \frac{m^2}{2}\chi-\frac{1}{2} j\frac{1}{-\partial^2+m^2}j \right)
 \nonumber  \\
& +\frac{1}{2}\ln\det \left[ \begin{matrix} \tilde V''(\chi)&  -i/2 \cr -i/2 & \Delta/2+{\mathcal J}
 \cr \end{matrix}\right]+\ldots
 \label{effectiveaction2}
\end{align}
where the ellipses stand for corrections of order 1/N and higher.
When we assume that the the source $j$ and the classical fields $m^2$ and $\chi$ 
are constants, we obtain the effective action evaluated on constant fields, 
\begin{align}
&  S=\text{N}\int\left\{-\frac{1}{12\pi}\left[m^2\right]^{\frac{3}{2}} 
+  V(\chi) - \frac{m^2}{2}\chi-\frac{j^2/\text{N}}{2m^2}  \right. 
 \nonumber  \\
& 
+\frac{1}{2\text{N}}\int \frac{d^{3}p}{(2\pi)^{3}}\ln\left[1+2V''(\chi)\left(\Delta(p)+\frac{2j^2/\text{N}}{m^4(p^2+m^2)}\right)\right]
 \nonumber \\ & \left. +\ldots   \right\} 
 \label{effectiveaction33}
\end{align}
Corrections represented by the ellipses in the last line of (\ref{effectiveaction33}) 
are functions of $1/\text{N}^2$ or higher order with $m^2,\chi$ and $j$ and terms with derivatives of $m^2,\chi$ and $j$. 

The first line in (\ref{effectiveaction33}) is the leading order in large N and the second line is the next-to-leading order.
 The integral in the next-to-leading order is ultra-violet divergent and  renormalization is required.  
The linear term in the effective action in a Taylor expansion in $j^2/\text{N}$ is 
 \begin{align}
 &-\frac{j^2}{2\text{N}}\left[\frac{1}{m^2}-\frac{4}{\text{N}}\frac{g^2\chi}{m^4}\int \frac{d^3p}{(2\pi)^3}
 \frac{1}{p^2+m^2}\frac{1}{1+2g^2\chi\Delta(p)} \right]\nonumber \\
&=- \frac{j^2}{2\text{N}}\left[\frac{1}{m^2}-  \frac{4}{\text{N}}\frac{g^2\chi}{m^4}\int \frac{d^{3}p}{(2\pi)^{3}} \frac{1-2g^2\chi\Delta(p)}{ (p^2+m^2)}  \right. \nonumber \\   
&~~~~~~~~~~\left. -\frac{4}{\text{N}}\frac{g^2\chi}{m^4}\int' \frac{d^3p}{(2\pi)^3}
 \frac{1}{p^2+m^2}\frac{1}{1+2g^2\chi\Delta(p)}\right]
\nonumber \\ \label{mass_renormalization}
&\equiv - \frac{j^2}{2\text{N}}\frac{1}{M^2} 
\end{align}
where the parameter $M$ is proportional to the renormalized mass of the $\vec\phi$-field.  At this order in the large N expansion,
the $\vec\phi$-field wave-function renormalization is finite.  The prime
on the integration in the third line means that the first two divergent terms which are
written before it have been subtracted, resulting in a finite integral. 
(These  divergent terms  are the first two terms in
a Taylor expansion of the integral in $g^2$.) Keeping $M$ 
finite as the ultra-violet cut-off is scaled to infinity requires  that we take $m^2$ to be a divergent function of $M^2$, 
 \begin{align}
m^2 =M^2 - \frac{4}{\text{N}} g^2\chi \left(
\frac{\Lambda}{2\pi^2}-\frac{M}{4\pi}\right) + \frac{g^4\chi^2}{2\pi^2\text{N}}\ln\frac{\Lambda}{\xi_1 M}
\nonumber \\
-\frac{4}{\text{N}} g^2\chi \int' \frac{d^3p}{(2\pi)^3}
 \frac{1}{p^2+M^2}\frac{1}{1+2g^2\chi\Delta(p)} \end{align}
where $\xi_1$ parameterizes the finite part of the logarithmically divergent integral.  
The effective action is
 \begin{align}
&  S  =\text{N}\int\left\{-\frac{1}{12\pi}M^3 
+  V(\chi) - \frac{M^2}{2}\chi-\frac{j^2/\text{N}}{2M^2}   \right.
 \nonumber  \\
& +\left( \frac{\chi}{2}+\frac{M}{8\pi}\right)\left[    \frac{4}{\text{N}} g^2\chi \left(
\frac{\Lambda}{2\pi^2}-\frac{M}{4\pi}\right) - \frac{g^4\chi^2}{2\pi^2\text{N}}\ln\frac{\Lambda}{\xi_1 M}
   \right.  \nonumber \\
&~~~~~~~~~~~~~~\left.  + \frac{4}{\text{N}} g^2\chi \int' \frac{d^3p}{(2\pi)^3}
 \frac{1}{p^2+M^2}\frac{1}{1 + 2g^2\chi\Delta(p)} \right]
\nonumber\\  & 
+\frac{1}{2\text{N}}\int' \frac{d^{3}p}{(2\pi)^{3}}{\ln}\left[1+2g^2\chi\left(\Delta(p)+\frac{2j^2/\text{N}}{M^4(p^2+M^2)}\right)\right]
\nonumber\\ &  \left. 
+\frac{1}{\text{N}}\int\frac{d^{3}p}{(2\pi)^{3}}\left[V''\Delta-\left( V''\Delta\right)^2 +\frac{4}{3}\left( V'' \Delta \right)^3\right]
+\ldots \right\}
\label{effectiveaction35}
\end{align}
As before, the prime on the integral in the fourth line indicates that the term of order $j^2/\text{N}$ and the divergent terms which are
written in the fifth line (these are the first, second and third order terms in a Taylor expansion in $g^2$) have been
 subtracted to render the integral finite. The terms that have been introduced by the mass renormalization
 are proportional to $\left(\frac{\chi}{2 }+\frac{M}{8\pi}\right)$ which vanishes on-shell. Here, we will first renormalize
 the effective action off-shell and then later on we will put the variables on-shell. We will find that the action is both on-shell and off-shell renormalziable. The divergent terms in the fifth line are
$$
\frac{1}{\text{N}}\int
\left[V''\Delta-\left( V''\Delta\right)^2 +\frac{4}{3}\left( V'' \Delta \right)^3\right]
=\frac{g^2\chi}{\text{N}}\left(\frac{\Lambda}{2\pi^2}-\frac{M}{4\pi}\right)^2 
$$
$$
-\frac{g^4\chi^2}{\text{N}}  \frac{\frac{\pi}{2} \Lambda - 4M\ln\frac{\Lambda}{M\xi_2}}{2^6\pi^3}
 +\frac{g^6\chi^3}{3\cdot 2^8\pi^2}\ln\frac{\Lambda}{M\xi_3}
$$
where $\xi_2$ and $\xi_3$ are constants which 
parameterize the finite parts of divergent integrals. 

Putting these in the effective action, we find a miraculous cancellation.  All divergent terms with a power of $M$
in the numerator cancel.  The remaining divergent terms can be canceled by counter-terms added to $V(\chi)$ alone.
What remains is    \begin{align}
&  S  =\text{N}\int\left\{-\frac{1}{12\pi}M^3 
+ \frac{g^2}{6}\chi^3 - \frac{M^2}{2}\chi-\frac{j^2/\text{N}}{2M^2}   \right. 
 \nonumber  \\
&   
 -\left(\chi+\frac{M}{4\pi}\right)     
 \frac{g^2\chi M}{2\pi \text{N}} +\frac{g^2\chi M^2}{16\pi^2\text{N}}   -   \frac{g^4\chi^2 M}{16\pi^3\text{N}}\ln\frac{\xi_2}{\xi_1}  \nonumber \\ 
 &  +
 \left(\chi+\frac{M}{4\pi}\right)\frac{g^2\chi M}{\pi^2\text{N}}\int' dp  
  \frac{p^2}{p^2+1}\frac{1}{1 + \frac{g^2\chi}{M}\frac{\arctan p/2 }{2\pi p} }  \
 \nonumber  \\
 & 
+\frac{M^3}{4\pi^2\text{N}}\int' dp p^2 \ln\left[1+\frac{g^2\chi}{M}\left(\frac{\arctan p/2 }{2\pi p}+\frac{4j^2/\text{N}M^6}{(p^2+1)}\right)\right]
\nonumber\\ &  \left. 
- \frac{g^4\chi^3}{4\pi^2\text{N}}\ln\frac{\mu}{\xi_1 M}
  +\frac{g^6\chi^3}{3\cdot 2^8\pi^2\text{N}}\ln\frac{\mu}{M\xi_3}+\ldots\right\}
\label{effectiveaction36}
\end{align}
We shall set $j^2=0$ and seek solutions of the equations of motion 
\begin{align}\label{eqmo_M}
&\frac{\delta S}{\delta M}=0
\\ \label{eqmo_chi}
&\frac{\delta S}{\delta \chi}=0
\end{align}
There are three important lessons to be learned from the form of the effective action (\ref{effectiveaction36}).  
\begin{enumerate}
\item{}First of all, to this order in 1/N the theory is off-shell renormalizable. 
The effective action that we have computed can be used to find the renormalized correlation functions of $\phi,\chi,im^2$-fields
where all external lines have vanishing momenta.  
\item{}The second lesson is that scale invariance is indeed violated at next-to-leading
order in large N, by the last two, logarithmic terms in (\ref{effectiveaction36}). From those terms we can find  the beta function for the
$\frac{g^2}{6\text{N}^2}(\vec\phi^2)^3$ interaction.  The effective action is a physical quantity, the volume times the energy of the theory when the 
fields are constrained to have certain expectation values.  As such, it should not depend on the renormalization scale $\mu$.  This is so 
if $g$ depends on $\mu$ in such a way that the action does not depend on $\mu$.  This yields
$$
\frac{\partial}{\partial\mu}\left( g^2(\mu)-\frac{1}{\text{N}}\left[\frac{3g^4(\mu) }{2\pi^2} 
  -\frac{g^6(\mu) }{ 2^7\pi^2}
\right]\ln\frac{\mu}{M}\right)=0
$$
\begin{align}
\beta (g)= \mu \frac{d}{d\mu}g^2(\mu)= \frac{1}{\text{N}}\left(  \frac{3g^4 }{2\pi^2} 
  -\frac{g^6 }{ 2^7\pi^2}\right)+\ldots
\end{align}
where the ellipses denote contributions of order 1/N and higher. 
This result matches the large N limit of the known perturbative beta-function \cite{Pisarski:1982vz}. 
\item{}The third important feature of the effective action in (\ref{effectiveaction36}) is that the argument of the logarithms in the $\mu$-dependent terms
contains only $M$ and $\mu$, and 
not $\chi$. Moreover, its coefficient contains only $\chi^3$ and does not depend on $M$.   
As a result of this structure, the equation of motion for $M$, (\ref{eqmo_M}), 
does not depend on $\mu$, and it is therefore scale invariant.  If we set $j^2=0$, $\frac{\delta}{\delta M}  S$  
is a homogeneous function of $\chi$ and $M$ and it is therefore  solved by $M=\alpha \chi$ whence it gives an equation for $\alpha$.
That equation is solved by $\alpha=-4\pi +\delta\alpha$ where $\delta\alpha\sim\frac{1}{\text{N}}$. 
\end{enumerate}

We can write the effective action in the form
\begin{align}
\frac{S}{\text{N}}= \int  \frac{\chi^3}{6 }\left[ 
    g^2(\mu) -g^{*2}(\tfrac{M}{\chi},g)    -  
   \beta(g(\mu) )  \ln\frac{\mu}{M} +\ldots \right]
\label{veff1}
\end{align}
where  
\begin{align}\label{offshellgstar}
g^{*2}\left(x,g\right)=\frac{1 }{ 2\pi  } x^3  
+3x^2+{\mathcal O}\left(\frac{1}{\text{N}}\right)
\end{align}
and the $g$-dependence is only in the higher orders in 1/N and 
can be substituted for its leading order $g=4\pi$. Also, on-shell, 
\begin{align}
g^{*2}\left(x_0,g\right)=(4\pi)^2+{\mathcal O}\left(\frac{1}{\text{N}}\right)  \label{onshellgstar}
\end{align}
In $-\frac{\chi^3}{6}g^{*2}$, we have gathered all of the terms in the effective action (\ref{effectiveaction36}) 
except those proportional to $\ln\frac{\mu}{M}$ in the last line and
the $\frac{g^3}{6}\chi^3$ term in the first line.  
In the last equality, we have used the leading order solution of the equation $\delta S/\delta M=0$, which is 
, $\frac{M}{\chi_0}=-4\pi+{\mathcal O}(1/\text{N})$ in $g^{*2}$.     

Then minimum of (\ref{veff1}) occurs at
\begin{align}\label{condensate}
M= \mu\exp\left(\frac{ g^{*2}-g^2 }{  \beta(g)}-\frac{1}{3}\right)
\end{align}
where we use equation (\ref{onshellgstar}) for $g^*$. This solution 
 is non-perturbative, both in the sense that, since $\beta\sim\frac{1}{\text{N}}$, 
it does not have a Taylor expansion in 1/N,  and in the sense that, when it is substituted into the 
effective action,   the logarithm produces a factor of $\frac{1}{\beta}\sim \text{N}$ which invalidates the large N expansion. 
In higher orders, powers of $\ln\frac{M}{\mu}$ will produce factors of N which can cancel their large N suppression.
This is similar to the phenomenon in the scalar field theory example in Coleman and Weinberg's work \cite{Coleman}  on 
dynamical symmetry breaking. There, they used the renormalization group to re-sum higher order logarithmic terms to obtain
a more accurate result. When they did, the extremum went away - there was no longer a symmetry breaking solution.  In the
present case, we will be more fortunate.  What allows us to find a solution is the presence of $g^*$ in the action. To begin,  
we will use the renormalization group to sum the leading logarithms of perturbation theory to all orders.   
 In this particular case, it is very simple. We  replace the combination which occurs in the effective action,  
\begin{align}\label{replace}
g^2(\mu)- \beta(g(\mu)) \ln\frac{\mu}{M}~, 
\end{align}
by the  running coupling at scale $M$, $g^2(M)$,  which is defined by integrating the beta function
\begin{align}
\int_{g^2(\mu)}^{g^2(M)} \frac{dg^2}{\beta(g)}=\ln\frac{M}{\mu}
\end{align}
The result of the integral, $g^2(M)$, has a 1/N expansion and 
the leading terms reproduce (\ref{replace}).  The corrections have higher
orders in $\frac{1}{\text{N}}\ln\frac{\mu}{M}$. The renormalization group improved potential energy of the effective action is then
the one given in equation (\ref{veff_1}), which we recopy here for the reader's convenience,
\begin{align}\nonumber
S =\text{N}\int d^3x\left\{ \frac{\chi^3(x)}{6}  \left(g^2(M(x))-g^{*2}\left(\tfrac{M(x)}{\chi(x)}\right)\right)+~~ \right. \\ \nonumber 
\left. +\frac{ \partial M(x)\cdot\partial M(x)}{96\pi | M(x) |}+\ldots\right\}
\end{align}
We will now study the states of the theory using this effective action.
The equations of motion are,
\begin{align} \label{em1}
&0=\frac{\delta S}{\delta \chi(x)} =  \frac{\chi_0^2}{2}\left(g^2(M_0)-g^{*2}\right) + \frac{ \chi_0M_0}{6}g^{*2\prime}
\\
&0=\frac{\delta S}{\delta M(x)} =  \frac{\chi_0^3}{6} \left(  \frac{1}{M_0}\beta -\frac{1}{\chi_0}g^{*2\prime}  \right) 
 \label{em2}
\end{align}where $g^{*2\prime}$ is a derivative of $g^{*2}$ by its argument $M/ \chi$. $M_0$ and $\chi_0$ are the solutions of these equations. 
We have dropped the derivative terms since we assume that the solutions  will be constant fields.

Equations (\ref{em1}) and (\ref{em2})  imply
\begin{align}\label{em3}
& g^{*2\prime} = \frac{\chi_0}{M_0}\beta(g^2(M_0))\\
 \label{em4}
& g^2(M_0) -  g^{*2}~=~-\frac{1}{3}\beta(g^2(M_0))
\end{align}
Equation (\ref{em3}) is an algebraic equation containing terms of order one and of order $\frac{1}{N}$ and
the variables  $\frac{M_0}{\chi_0}$ and $g^2$. $g^2$ appears only in the terms of order $\frac{1}{N}$ and it can
therefore be regarded as a constant, and set to $4\pi$. 
In the leading order, equation (\ref{em3}) has the solution
$\frac{M_0}{\chi_0}=-4\pi+{\mathcal O}(1/\text{N})$ and the order $\frac{1}{N}$ terms are easily computable. 

We recall that $g^{*2}$ has a similar structure to equation (\ref{em3}), it contains terms of order one and of order $\frac{1}{N}$ and
the variables  $\frac{M_0}{\chi_0}$ and $g^2$,  and  $g^2$ appears only in the terms of order $\frac{1}{N}$. 
We can then plug the solution for $\frac{M_0}{\chi_0}$ which we discussed in the above paragraph into $g^*$ to obtain
a $\frac{1}{N}$ corrected expression for it.

Then we use the corrected $g^{*2}$ in equation (\ref{em4}).  The solution of equation (\ref{em4}) is a mass scale, that is, 
the value of the mass scale where the running coupling solves the equation.  This yields the value of the condensate $M_0$ and
the above considerations then determine $\chi_0=-\frac{1}{4\pi}M_0+{\mathcal O}(\frac{1}{N})$.    Due to the order $\frac{1}{N}$
violation of scale invariance, and unlike the scale invariant infinite $N$ limit, 
the values of these condensates are no longer arbitrary, but they are fixed by the value
of the running coupling constant at some reference scale. 
     
We substitute the solution into the effective action and then obtain
\begin{align}
S_{\rm on-shell} =\text{N}\int\left\{  \frac{ M_0^3 }{18(4\pi)^3}  \beta(g( M_0)) +\ldots\right\}
\end{align}
where the ellipses are terms of order $\frac{1}{\text{N}^2}$ and higher.

To examine the fate of the dilaton,  we return to the action (\ref{veff_1}) and we
consider the fluctuation matrix about the solution that we have found,
\begin{align}
& \frac{1}{\text N}\frac{\delta^2 S }{\delta\chi^2}=\frac{\chi_0}{3}\beta - \frac{M_0^2}{6\chi_0}g^{*2\prime\prime}
\\
& \frac{1}{\text N}\frac{\delta^2 S}{\delta\chi\partial M} = \frac{\chi_0^2}{6M }\beta+\frac{M_0}{6}g^{*2\prime\prime}
\\
 & \frac{1}{\text N}\frac{\delta^2 S}{\delta M^2}= -\frac{\chi_0^3}{6M_0^2}\beta-\frac{\chi_0}{6} g^{*2\prime\prime} +\frac{p^2}{32\pi|M_0|}
\end{align}
where we have used equations (\ref{em3}) and (\ref{em4}) to simplify the right-hand-sides.

We can determine determinant of the fluctuation matrix and find,
\begin{align} \frac{1}{\text N^2}\det 
\left[ \begin{matrix}   \frac{\delta^2 S }{\delta\chi^2}&   \frac{\delta^2 S}{\delta\chi\partial M}
\cr  \frac{\delta^2 S}{\delta\chi\partial M} & \frac{\delta^2 S}{\delta M^2}
 \cr \end{matrix}\right] = \frac{M_0^2}{32\pi^2}\beta-\frac{p^2}{4}
\label{det} \end{align}
where we have used the fact that,
 \begin{align}
 g^{*2}&= \frac{1}{2\pi} \frac{M^3}{\chi^3}+\frac{3M^2}{\chi^2}+ \ldots  =(4\pi)^2+ \ldots 
\end{align}
 and,
 \begin{align}
 g^{*2\prime\prime}&= \frac{3}{\pi}\frac{M}{\chi}+6+\ldots  
=-6+\ldots
\end{align}

The determinant of the the fluctuation matrix is proportional to  the inverse propagator of the $\chi$- and $M$- fields. 
The beta function is positive over the interesting range of $g^2$ (see FIG. \ref{beta}).
Clearly, from equation (\ref{det}), we see that these excitations are tachyonic  with mass given by,
\begin{align}\label{mass}
m^2_\text{dilaton}=-\frac{3M_0^2}{8\pi^2}\beta .
 \end{align}

We conclude our paper by summarizing our results. We found that, although at leading order in 1/N, phi-six theory in three dimensions exhibits spontaneous scale symmetry breaking accompanied by a massless dilaton, at the next-to-leading order in 1/N,  dilaton acquires a tachyonic mass and   the spontaneously broken phase is
therefore unstable. Our background field technique  found the perturbative beta function of the O(N) symmetric $g^2(\phi^2)^3$  theory. The result agreed with
the beta function originally found by Pisarski \cite{Pisarski:1982vz}.  
Our result is that the tri-critical behaviour which is described by the massless O(N) symmetric $g^2(\phi^2)^3$ theory is stable over a 
larger range of coupling constants that it is commonly thought to be. Of course, our analysis applies only 
if $N$ is is not infinite, but if it is large enough that our large $N$ expansion is accurate.
 
\noindent
We thank Bill Bardeen, Moshe Moshe and Rob Pisarski for comments. The work of H.O.~and G.W.S.~ was supported by NSERC of Canada.  
Research of L.C.R.W. was partially supported by a University of Cincinnati Faculty Development Grant.  He thanks Mainz Center for 
Theoretical Physics for their hospitality during the 2016 Workshop on Composite Dynamics, and conversations with B.Holdom,  
V.Miransky, E.Pallante  and F.Sannino.

\end{document}